\documentclass[pra,aps,twocolumn,showpacs,superscriptaddress,nofootinbib,floatfix]{revtex4-1}
\usepackage[colorlinks=true,citecolor=blue,linkcolor=magenta]{hyperref}
\usepackage{graphicx,stackrel}
\usepackage{dcolumn}
\usepackage{bm}
\usepackage[bbgreekl]{mathbbol}
\usepackage{amsmath,dsfont}
\usepackage[normalem]{ulem}
\usepackage{color}
\usepackage[ruled,vlined]{algorithm2e}
\usepackage[utf8]{inputenc}

\usepackage{booktabs}

\newcommand{\bigO}{\mathcal{O}}

\newtheorem{theorem}{Theorem}

\begin{document}
\title{Exploring the impact of graph locality for the resolution of MIS with neutral atom devices}
\author{Constantin Dalyac}
\affiliation{PASQAL, 7 rue Léonard de Vinci, 91300 Massy}
\affiliation{LIP6, CNRS, Sorbonne Université, 4 Place Jussieu, 75005 Paris, France}

\author{Louis-Paul Henry}
\affiliation{PASQAL, 7 rue Léonard de Vinci, 91300 Massy}

\author{Minhyuk Kim}
\affiliation{Department of Physics, Korea Advanced Science and Technology (KAIST), Daejeon 34141, Republic of Korea}

\author{Jaewook Ahn}
\affiliation{Department of Physics, Korea Advanced Science and Technology (KAIST), Daejeon 34141, Republic of Korea}

\author{Loïc Henriet}
\affiliation{PASQAL, 7 rue Léonard de Vinci, 91300 Massy}

\date{\today}
\begin{abstract}
In the past years, many quantum algorithms have been proposed to tackle hard combinatorial problems. In particular, the Maximum Independent Set (MIS) is a known NP-hard problem that can be naturally encoded in Rydberg atom arrays. By representing a graph with an ensemble of neutral atoms one can leverage Rydberg dynamics to naturally encode the constraints and the solution to MIS. However, the classes of graphs that can be directly mapped ``vertex-to-atom" on standard devices with 2D capabilities are currently limited to Unit-Disk graphs. In this setting, the inherent spatial locality of the graphs can be leveraged by classical polynomial-time approximation schemes (PTAS) that guarantee an $\epsilon$-approximate solution. In this work, we build upon recent progress made for using 3D arrangements of atoms to embed more complex classes of graphs. We report experimental and theoretical results which represent important steps towards tackling combinatorial tasks on quantum computers for which no classical efficient $\varepsilon$-approximation scheme exists.
\end{abstract}

\maketitle



In the past decade, Quantum Processing Units (QPUs) consisting of atoms trapped in arrays of optical tweezers have been used extensively to address quantum simulation problems with spin systems\,\cite{Browaeys2020,henriet_quantum_2020}. The techniques and methods developed can also be leveraged to explore the resolution of combinatorial optimization problems, as all of Karp’s 21 NP-complete problems can be reformulated as ground-states of Ising models\,\cite{Lucas_2014}. The authors of the pioneer publication\,\cite{Pichler2018} noticed that the Hamiltonian of interacting Rydberg atoms naturally realizes the cost function of the Maximum Independent Set (MIS) problem on the graph induced by the atoms in interaction. This feature enables the implementation of adiabatic or variational schemes such as the QAA\,\cite{Farhi_adiabatic} or the QAOA\,\cite{Farhi2014} algorithms to approximately solve the MIS problem. The set of graphs that can be solved corresponds to Unit Disk (UD) graphs, where vertices are represented as points in the Euclidean plane and two vertices are connected by an edge if the distance between the two corresponding points is lower than a threshold value.

A recent implementation on more than 280 atoms has attracted a lot of attention\,\cite{Ebadi2022}, with the observation of heuristic resolution of the MIS problem on a large set of UD graphs. To grasp the performances of such quantum approximation algorithms, we need to compare them to their classical counterparts. In the case of the MIS problem on UD graphs, there exist efficient classical approximation algorithms with guaranteed performance ratios called Polynomial-Time Approximation Schemes (PTAS). In Section\,\ref{sec:UD-ptas}, we analyze the effects of graph locality on the approximability of the solutions to the problem, illustrating that the presence of structure (locality, planarity) in a graph can provide enough information for a classical algorithm to find good approximations in polynomial time, which leaves little room for quantum advantage perspectives in those cases. In order to leave the classes of efficient classical approximation schemes, several strategies can be adopted and/or combined, including the incorporation of ancillary vertices\,\cite{Pichler2018,Nguyen22,Lanthaler23}. In this paper, we use in addition the capabilities of neutral atom devices to solve the MIS problem in 3D\,\cite{Byun22,Kim22}. In Section\,\ref{sec:Unit-ball_exp}, we start to expand the class of graphs on which one can approximately solve MIS by considering $K_{33}^+$, a non-UD graph with 7 vertices in 3D. This graph is the smallest graph known for which the optimal classical greedy algorithm fails most of the time. In Section \ref{sec:embedding}, we then present a systematic and efficient method to map a MIS of any general graph of max-degree $6$ onto the ground state of an ensemble of interacting neutral atoms in 3D. For such graphs, no classical algorithm is known to find an $\varepsilon-$approximate solution in polynomial time.  \\

\section{Solving MIS on local graphs}
\label{sec:MIS}

Given a graph $G=(V, E)$, an independent set is defined as a subset $S$ of the vertices such that no two vertices of $S$ share an edge in $G$. Mathematically, $S$ is an independent set of $G$ \textit{iff} $S \subseteq V / \forall (x,y) \in S^2 , (x,y) \notin E$. A maximum independent set $S^*$ corresponds to an independent set of maximum cardinality. 

Any possible solution to this problem consists in separating the vertices of $G$ into two distinct classes: an independent one and the others. We attribute a status $z$ to each vertex, where $z_i = 1$  if vertex $i$ belongs to the independent set, and $z_i=0$ otherwise. The Maximum Independent Sets correspond to the minima of the following cost function: 

\begin{equation}
   C(z_1,\dots,z_N) = -\sum_{i=1}^N z_i + U \sum_{\langle i,j \rangle}z_i z_j
 \label{cost_function}
\end{equation}
where $U \gg \Delta(G)$, and $\Delta(G)$ is the degree of the vertex with  maximum degree, $\langle i,j \rangle$ represents nodes in $E$ (the edges), and $N=|V|$. This cost function favours having a maximal number of atoms in the $1$ state, but the fact that $U \gg 1$  strongly penalizes two adjacent vertices in state 1.

Interestingly, the cost function of Eq. (\ref{cost_function}) can be natively realized on a neutral atom platform\,\cite{Pichler2018}, with some constraints on the graph edges. Placing $N$ atoms at positions $\textbf{r}_j$ in a 2D plane, and coupling the ground state $|0\rangle$ to the Rydberg state $|1\rangle$ with a laser system enables the realization of the Hamiltonian :
\begin{equation}
    H= \sum_{i=1}^N \frac{\hbar\Omega}{2} \sigma_i^x - \sum_{i=1}^N \frac{\hbar \delta}{2}  \sigma_i^z+\sum_{j<i}\frac{C_6}{|\textbf{r}_i-\textbf{r}_j|^{6}} n_i n_j.
\label{eq:ising_Hamiltonian}
\end{equation}
Here, $\Omega$ and $\delta$ are respectively the Rabi frequency and detuning of the laser system and $\hbar$ is the reduced Planck constant. The first two terms of Eq. (\ref{eq:ising_Hamiltonian}) govern the transition between states $|0\rangle$ and $|1 \rangle$ induced by the laser, while the third term represents the repulsive van der Waals interaction between atoms in the $|1\rangle$ state. More precisely, $n_i = (\sigma_i
^z + 1)/2$ counts the number of Rydberg excitations at position $i$. The interaction strength between two atoms decays as $|\textbf{r}_i-\textbf{r}_j|^{-6}$ and $C_6$ is a constant which depends on the chosen Rydberg level. 

The shift in energy originating from the presence of two nearby excited atoms induces the so-called \textit{Rydberg blockade} phenomenon. More precisely, if two atoms are separated by a distance smaller than the Rydberg blockade radius $r_b = (C_6/\hbar \Omega)^{1/6}$, the repulsive interaction will prevent them from being excited at the same time. On the other hand, the sharp decay of the interaction allows us to neglect this interaction term for atoms distant of more than $r_b$. As such, for $\Omega=0$, the Hamiltonian in Eq. (\ref{eq:ising_Hamiltonian}) is diagonal in the computational basis and enables to realize $H |z_1,\dots,z_N\rangle=(\hbar \delta/2) C(z_1,\dots,z_N)|z_1,\dots,z_N\rangle$, with the cost function specified in Eq. (\ref{cost_function}), and for which there is a link between atoms $i$ and $j$ if they are closer than $r_b$ apart.\\


\section{Local geometric structure implies PTAS}
\label{sec:UD-ptas}

\begin{figure}[ht]
    \centering
    \includegraphics[width =1.\linewidth]{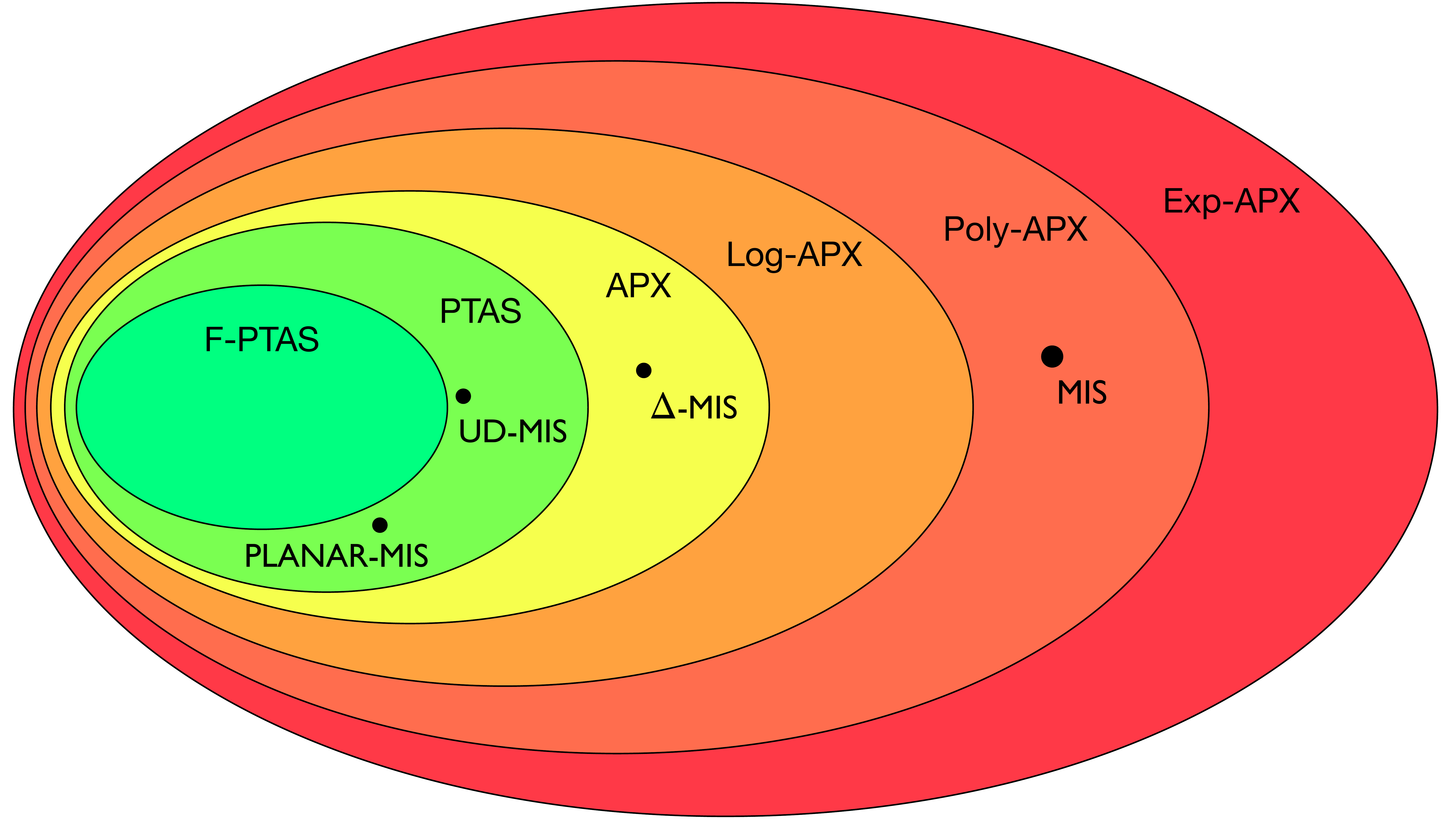}
    \caption{\textbf{Standard approximability classes of NP-hard problems} (under the assumption $P \neq NP$). While solving MIS exactly is NP-Hard, finding an approximate solution of useful quality can be easy, depending on the nature of the graph at hand. The complexity class NPO corresponds to the class of optimisation problems whose underlying decision problem is in NP. A polynomial-time approximation scheme (PTAS) is a family of $\varepsilon-$parameterised algorithms that output approximate optimal solutions that are $\varepsilon > 0$ away from the exact solution in polynomial time. When the family of algorithms is also polynomial in the parameter $\epsilon$, it is a Full-PTAS (F-PTAS) and is very efficient. The APX class corresponds to problems for which a polynomial-time algorithm can only achieve a constant approximation ratio. For MIS, if the graph is Unit-Disk or planar, there exists a PTAS. In our work, we propose a polynomial embedding of general bounded-MIS ($\Delta$-MIS) problems which are known to be in APX. $\Delta$-MIS is even APX-Complete, meaning that no PTAS exists for it unless $P = NP$. In the case of a general graph with no underlying structure, it was proven that for any $\varepsilon > 0$, MIS is inapproximable within approximation ratio $n^{\varepsilon - 1}$, corresponding to the complexity class poly-APX. }
    \label{fig:Approx_classes}
\end{figure}

Unit-Disk graphs are inherently local in the sense that two vertices $v$ and $w$ are connected by an edge if and only if the distance between the two vertices is inferior to a given threshold. While finding the exact solution remains $NP-$hard, there exists efficient approximations to the solution. Known results about approximations to the general MIS problem are presented in Figure \ref{fig:Approx_classes}. Interestingly, the quality of the approximation depends on the type of graph under study: for Unit-Disk graphs, classical algorithms can leverage the locality of the edges to efficiently estimate an approximation of the MIS. The main idea is to split the graph into local subgraphs for which MIS is solved exactly. Aggregating the solutions of the subgraphs yields a good solution as a subgraph only affects its neighbouring subgraphs. It was shown that this method corresponds to a Polynomial-Time Approximation Scheme (PTAS) that guarantees a $1-\varepsilon$ approximation ratio in polynomial time\,\cite{hunt1998nc}. In the case of planar graphs, another PTAS exists that also guarantees high approximation ratios. In the same flavour as for the Unit-Disk case, it relies on dividing the graph into subgraphs with $k-$outerplanar forms \,\cite{Baker1994}.

\begin{figure*}[t!]
    \centering
    \includegraphics[width =\linewidth]{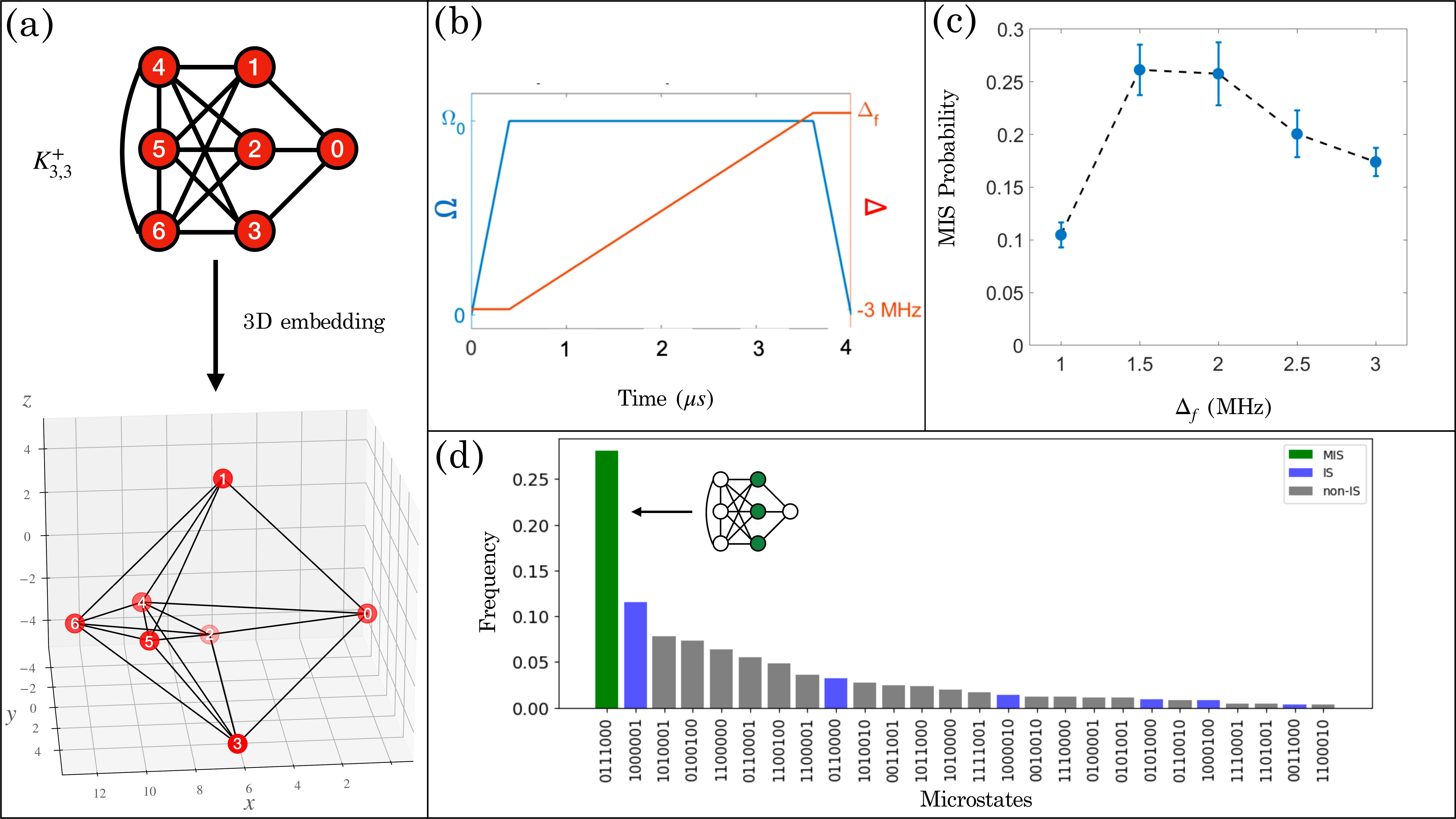}
    \caption{\textbf{Solving MIS on unit-ball graphs with Rydberg atoms.} The graph  $K_{3,3}^+$ represented in \textbf{(a)} is the smallest example of a unit-ball graph for which the Greedy algorithm fails. This graph is embedded with a 3D array of atoms. \textbf{(b.)} The adiabatic evolution path corresponds to a linear detuning $\Delta$ from $\Delta_i=2 \pi \times 0.7$MHz to $\Delta_f$, while the Rabi frequency $\Omega$ is ramps up to $\Omega_0= 2 \pi \times 0.7$MHz. \textbf{(c.)} The experiment is tested by changing the final detuning $\Delta_f$.  \textbf{(d.)}The microstates probability histogram is presented for $\Delta_f = 2 \pi \times 1.5$MHz. The SPAM error $P(g|r)=0.15$ and $P(r|g)=0.05$ is corrected following the method proposed in\,\cite{geller2020efficient}. The MIS solution probability is improved up to $P_{\text{MIS}} = 0.25$, in stark contrast to the other microstates.}
    \label{fig:K_33+}
\end{figure*}

However, more general graphs such as bounded-degree graphs do not present enough structure for classical algorithms to $\varepsilon-$approximate maximum independent sets in polynomial time. In the case of a graph with bounded-degree $\Delta$, finding an approximation solution to the MIS problem is known to be APX-complete\,\cite{Papadimitriou1991}. In other words, this means that the best approximation ratio that can be guaranteed by a polynomial-time classical algorithm is constant; to the best of our knowledge this approximation ratio is $r = \frac{5}{\Delta+3}$\,\cite{Berman1994}. This ratio cannot be improved without adding an exponential time overhead for a classical algorithm. The approximation guaranteed by a polynomial-time approximation scheme worsens in the case of a general graph. It was proven that for any $\varepsilon > 0$, MIS is inapproximable within approximation ratio $n^{\varepsilon - 1}$ unless $P=NP$\,\cite{zuckerman2006linear}. A summary of the approximability classes of MIS on these specific classes of graphs is shown in Fig. \ref{fig:Approx_classes}. The key take-away is that the presence of structure (locality, planarity) in a graph can provide enough information for a classical algorithm to find good approximations in polynomial time.

The difference in the approximability of these problems motivates the need for efficient hardware embedding of graphs which present less geometrical structure than UD or planar graphs. Unit-disk graphs can be extended to higher dimension where they are referred to as unit-ball graphs of dimension $d$. It can be shown that any $n-$vertex graph can be embedded as a unit-ball graph of dimension $d=n-1$\cite{MAEHARA198455}. The dimension of the embedding is therefore a parameter of hardness for MIS approximation, since for $d=2$ there exists a PTAS but for $d=n-1$ it is inapproximable within approximation ratio $n^{\varepsilon - 1}$ for any $\varepsilon > 0$. A natural step in finding a hard approximation zone is therefore to increase the dimension $d$, which we do in the following section by embedding unit-ball graphs where $d=3$.


\section{Experimentally solving MIS on unit-ball graphs in 3D arrays of atoms}
\label{sec:Unit-ball_exp}

A common strategy employed to find an approximate solution to the MIS problem is to use a greedy algorithm. This iterative method involves local optimal choices at each step and yields an acceptable solution in polynomial time. For example, a greedy algorithm on MIS selects a random vertex in $G$ at each step, deletes its neighbours and repeats this step until the graph is empty. By construction this method guarantees that the selected vertices form an independent set. The greedy algorithm therefore provides an $1/n-$approximation to MIS as in the worst-case it returns a single vertex. Surprisingly, the inapproximability result for general MIS below $n^{\varepsilon-1}$ implies that the greedy algorithm is an optimal approximation algorithm\,\cite{Hastad99}. It is also proven to be optimal for bounded-degree graphs\,\cite{Halldorsson97}. An improvement to the greedy algorithm is to select at each step the vertex with the lowest degree as it deletes less neighbouring vertices. As the greedy algorithm runs in polynomial time and MIS is an NP-complete problem, there exists a class of graphs for which the greedy algorithm fails completely. The smallest graph of this class\,\cite{Mari17} is $K_{3,3}^+$ and is represented in Fig.\ref{fig:K_33+}. It corresponds to the bipartite graph $K_{3,3}$ augmented with edges such that one class forms a clique (in our figure it corresponds to vertices $(4, 5, 6)$) and a single vertex (0) is connected to the other class $(1,2,3)$. In this case, the degree-informed greedy algorithms fails. Interestingly, $K_{3,3}^+$ is a unit-ball graph, for which we give an embedding on the right of  Fig.\ref{fig:K_33+}. We therefore embed $K_{3,3}^+$ experimentally in 3D using an array of atoms and follow a quantum annealing scheme to prepare its MIS.

The first trial adiabatic evolution path is tuning the detuning $\Delta$ linearly from $\Delta_i$ to $\Delta_f$, while the Rabi frequency $\Omega$ is fixed to $\Omega_0$. The experiment is tested by changing the final detuning $\Delta_f$. The full 27 microstates probability histogram is presented for $\Delta_f = 2 \pi \times 1.5MHz$. The SPAM error $P(g|r)=0.15$ and $P(r|g)=0.05$ is corrected. The MIS solution probability is improved up to $P_{\text{MIS}} = 0.25$, in stark contrast with the other microstates. These very encouraging results show that it is possible to prepare the MIS of $K_{33}^+$ using a quantum annealing scheme on a 3D array of neutral atoms. On the classical side, $K_{33}^+$ is the smallest example for which a greedy algorithm, optimal in approximation, fails. Other classical methods however known as \textit{slicing}\,\cite{van2006better} can return the optimal solution in polynomial time on such instances. There are classes of graphs however for which the inapproximability is stronger. This is the case for bounded-degree graphs and we will demonstrate how such graphs can be embedded on a neutral atom device in the following section.

\begin{figure*}[t!]
    \centering
    \includegraphics[width =0.8\linewidth]{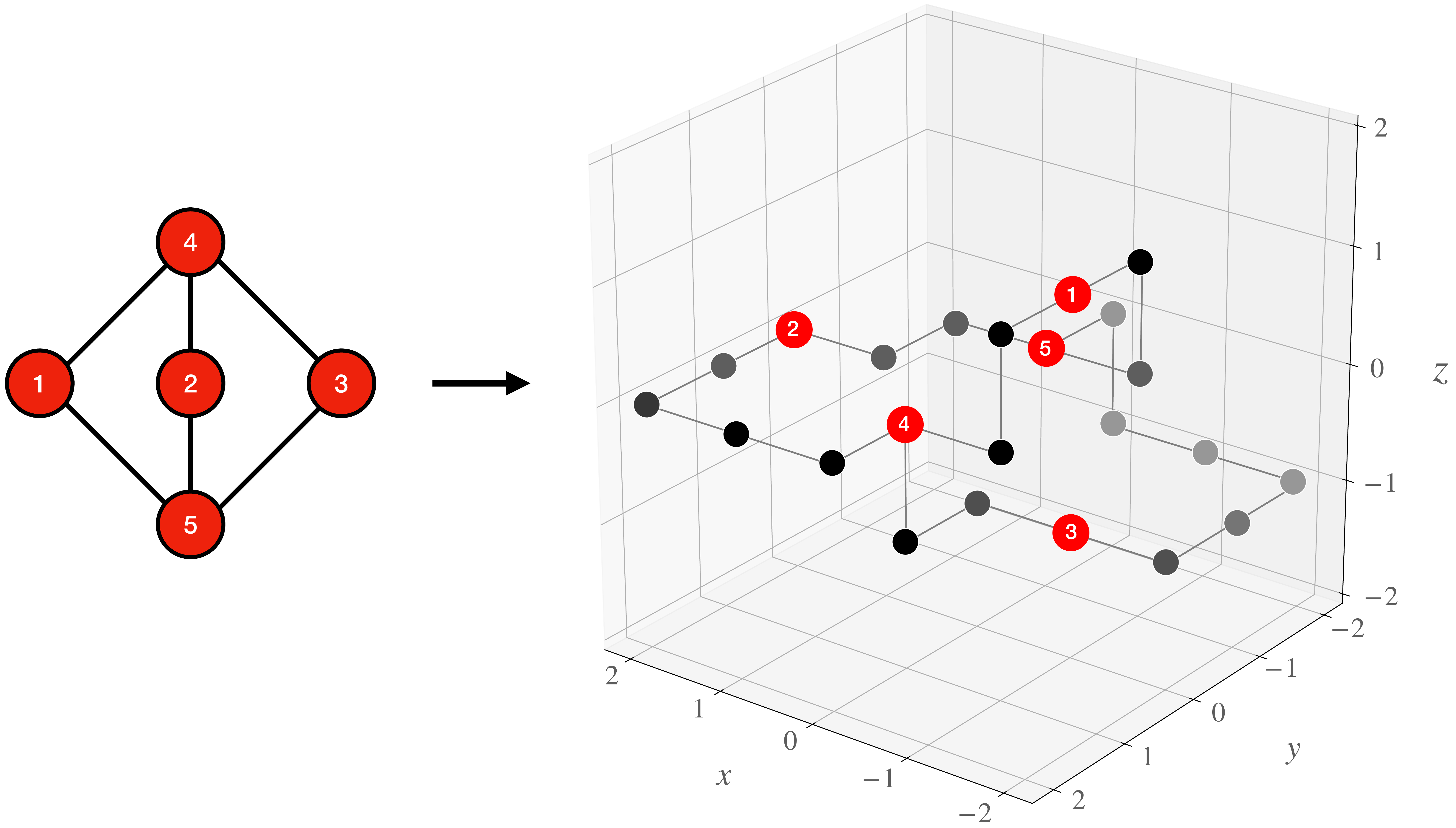}
    \caption{\textbf{Mapping a non-UD graph with Rydberg atoms using 3D quantum wires.}  The graph on the left is the smallest example of a non-Unit-Disk graph (otherwise the vertex 2 would be connected to 1 and 3). This graph is embedded in a 3D array of atoms (right). The red atoms correspond to the original graph vertices, while the chains of black atoms encode the edges of the original graph. The procedure enables to map any general graph, including non-UD and non-planar graphs with a bounded-degree up to $\Delta=6$.}
    \label{fig:3D_mapping}
\end{figure*}

\section{Embedding bounded degree graphs in 3D arrays of atoms}
\label{sec:embedding}


Recent and previous works propose to represent non-local edges of graphs with chains of ancillary atoms\,\cite{Pichler2018, Kim22}, in 2D and 3D respectively. Building upon this idea, we present an efficient and systematic method to represent any graph of degree inferior or equal to 6 with a 3D array of atoms. An illustration of our method is given in Figure \ref{fig:3D_mapping}. It runs in polynomial time and numerical simulations suggest a low overhead in the number of added ancillary atoms (sub-linear). \\


We define the \textit{drawing} of a graph as the realisation or layout of a graph in a 3D space, where no two vertices overlap and no vertex-edge intersection occurs unless its incidence exists in the original graph. We say that a drawing is \textit{crossing-free} if no two edges cross. A growing interest emerged in 3D drawings of graphs for circuit designs\,\cite{leighton1986three} or for information visualisation\,\cite{ware1994viewing, Ware08}. In our case, we focus on 3D orthogonal grid drawings (OGD) of a graph $G=(V,E)$ for which the vertices of $G$ are represented as distinct points of the grid $\mathbb{Z}^3$, while all edges $E={(u,v) \in V^2}$ are restricted to being drawn on lines parallel to one of the three axes. This restriction implies that only graphs with maximum degree six can have such a drawing; given a vertex at coordinates $(x,y,z) \in \mathbb{Z}^3$, the only authorised directions for an edge are $(x \pm 1,y \pm 1,z \pm 1)$. It is proven that every graph of bounded-degree $\Delta$ admits a crossing-free OGD if $\Delta \leq 6$\,: 

\begin{theorem}[\cite{Eades1996}]
    Let $G$ be a graph with $n$ vertices and maximum degree $\Delta \leq 6$. Then $G$ has a three-dimensional orthogonal grid drawing such that no pair of edges cross.
\end{theorem}

The general idea behind our method is to construct an OGD for the original graph and replace the edges by chains of ancillary atoms. The advantage of an OGD is that two distinct edges only intersect at common endpoints, thereby preventing ancillary atoms from interacting if they are not part of the same chain. Ideally, we would like to find an OGD that minimises the edge lengths in order to have as little ancillary atoms as possible. Unsurprisingly, it is $NP$-hard to find an OGD that minimises the total length of the edges\,\cite{Eades1996}. While many different algorithms have been proposed to optimise the total volume\,\cite{Biedl02orthogonaldrawings} or the average number of bends per edge\,\cite{Papakostas97} of an OGD, we present a simple heuristic to construct an OGD with a small total length of edges.

Given a general $\Delta$-bounded graph $G=(V, E)$ with $\Delta \leq 6$, we place the vertices of $V$ in $\mathbb{R}^3$ using the Fruchterman-Reingold algorithm (FR)\,\cite{fruchterman1991graph} that runs efficiently in $\bigO(|V|^3)$ steps. Note that other algorithms could be used at this step but FR yielded the best results in our simulations. The vertices are then moved to the closest grid point in $\mathbb{Z}^3$, insuring that no two vertices get the same coordinates.

We then use optimal path-finding algorithms\,\cite{Dijkstra1959,Hart1968} to find the shortest route between two vertices, restricted to the underlying grid. The resulting path is transformed into a chain of ancillary vertices. We previously ensure that two distinct edges are separated by at least a 2-grid-point distance, in order to avoid any unwanted interaction between ancillary atoms of two distinct chains. Finally, the parity of the length of the path has to be checked, in order to ensure it conveys the proper independent-set constraint. If the path length is odd, we add an ancillary vertex at each $\mathbb{Z}^3$ coordinate of the path. If the path is of even length $p$, we add $p+1$ evenly spaced ancillary vertices along the path. After this procedure, one obtains an augmented graph $G_+=(V_+, E_+)$ of size $|V_+| = N_+$. \\



By representing all vertices with atoms, one can encode a Maximum Independent Set of $G$ in the ground state of an Ising Hamiltonian on Rydberg atoms\,\cite{Browaeys16} over the augmented graph $G_+$ :

\begin{equation}
H_+=\sum_{j=1}^{N_+}\frac{\hbar\Omega}{2}  \sigma_j^x- \sum_{j=1}^{N_+} \frac{\hbar}{2}(\delta + \delta_j)\sigma_j^z+\sum_{j<i}\frac{C_6}{|\textbf{r}_i-\textbf{r}_j|^{6}} n_i n_j.
\label{ising}
\end{equation} where $n_i=(\sigma^z_i + \mathbb{1})/2$ and $\delta_j$ represents the local detuning applied to each atom, ensuring that the MIS is achieved when the chain is in an antiferromagnetic state.

\begin{figure}
    \centering
    \includegraphics[width=\linewidth]{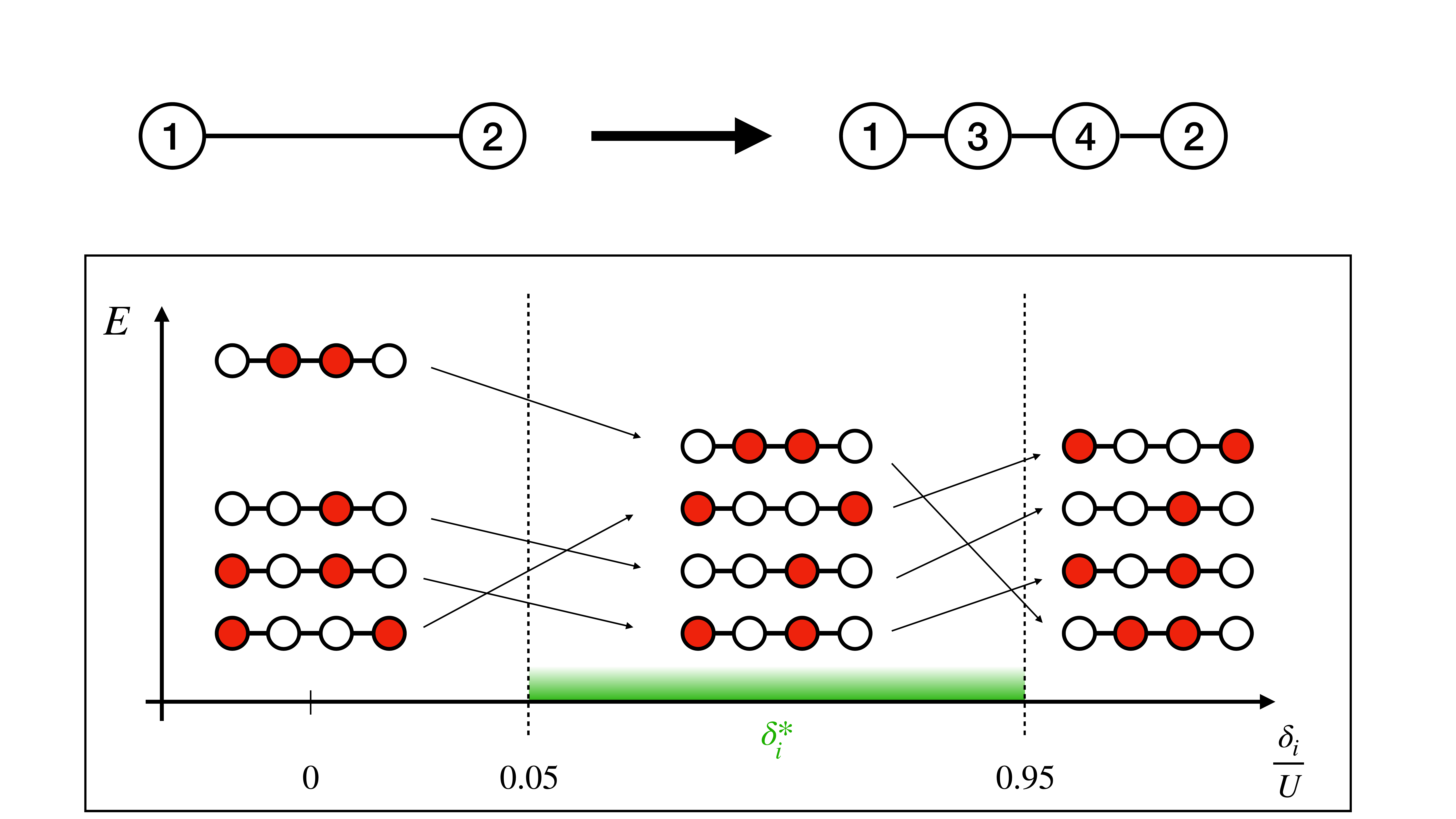}
    \caption{\textbf{Local detuning on ancillary atoms.} In the top, an edge between vertex 1 and 2 is augmented with two ancillary atoms 3 and 4. With no local detuning, the ground-state of the augmented edge corresponds to exciting $\{1,2\}$. This however is inconsistent in the initial graph (1 and 2 cannot be simultaneously in the MIS). We therefore add local detuning to all ancillary atoms to ensure that the ground-state corresponds to the anti-ferromagnetic state.  (b) Energy diagram of the spectrum as a function of the local detuning $\delta_i$ applied to the ancillary atoms. A red atom corresponds to an excited state. The independence condition of the original link is respected if $0.05 \times U \leq \delta_i \leq 0.95 \times U$, where $U$ is the interaction between two neighbouring atoms. In these values of local detuning, the ground-state corresponds to the anti-ferromagnetic state.}
    \label{fig:Augmented_path}
\end{figure}

In Figure \ref{fig:Augmented_path}, we show explicitly on a single augmented edge how one can choose the values of local detuning on ancillary atoms to ensure that the MIS of the edge corresponds to the groundstate of the Hamiltonian $H_+$.

In this example, there are 3 Maximum Independent Sets in the augmented graph : $\{1,4\},\{2,3\}$ which are acceptable solutions, but also $\{1,2\}$ which does not correspond to a MIS of the original graph. With a reasonable global detuning $\delta>0$, this latter state is actually the ground state of the chain. In order to guarantee that $\{1,4\}, \{2,3\}$ correspond to the ground-states of the Hamiltonian associated to the augmented path, we apply an additional local detuning $\delta_i$ to each ancillary atom with $\delta_i = J/2$, where $U=C_6/r^6$ is the interaction energy between two closest atoms of the augmented graph. 

Keeping $0.5 \times U \leq \delta_i \leq 0.95 \times U$ ensures that the MIS returned by the algorithm preserves initial constraints (the proof is given in appendix \ref{appendix_local_detuning}). In the general case, we determine for each edge the corresponding value for the detuning that will be applied on all the ancillary atoms of that chain. This ensures that the ground state of the augmented Hamiltonian encodes a Maximum Independent Set of the original graph.

\begin{figure}[ht]
    \centering
    \includegraphics[width =1.\linewidth]{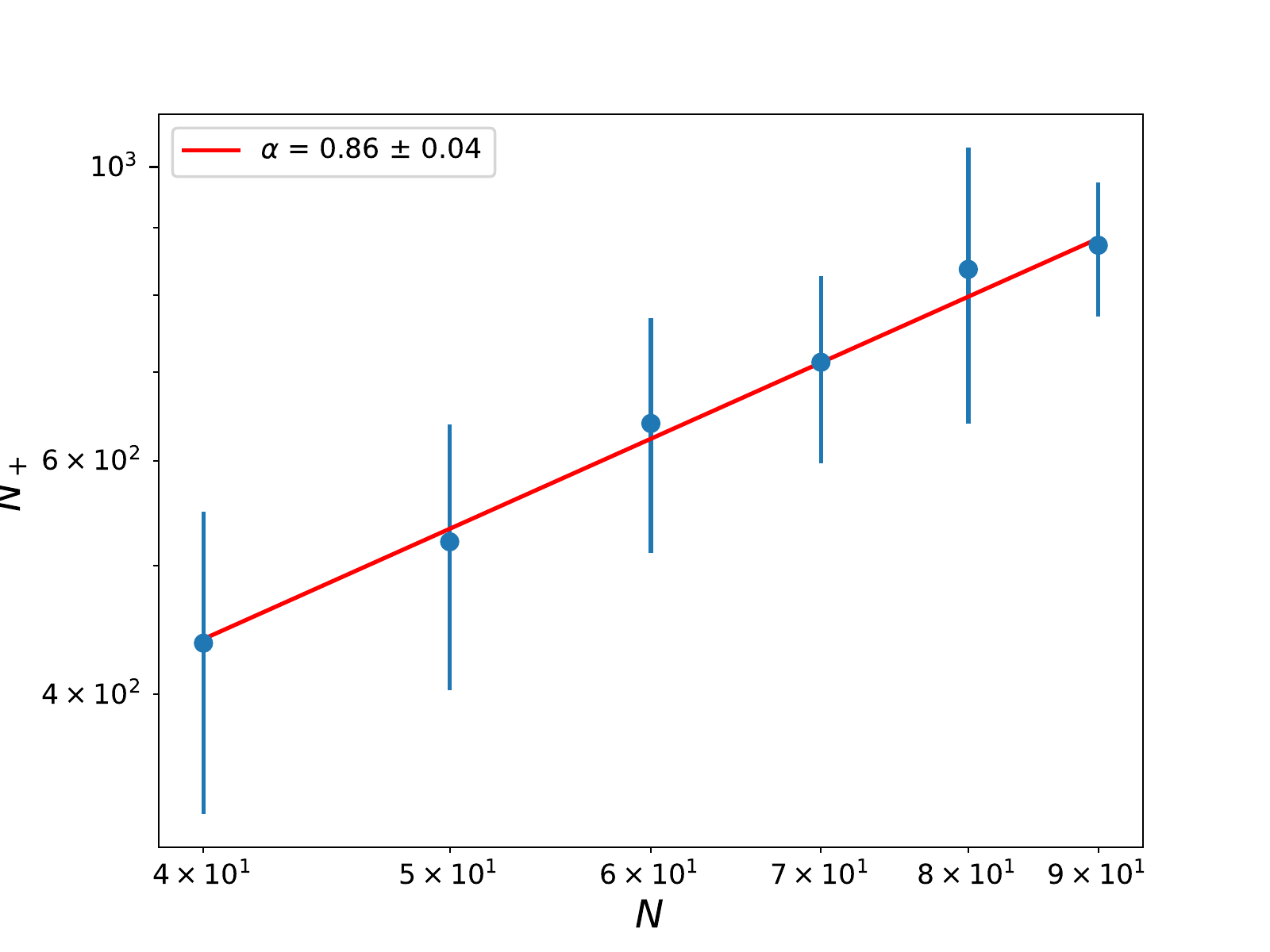}
    \caption{\textbf{Scaling of additional atoms in our method}. For each size, 20 random Erdos-Renyi graphs of maximum degree 6 are generated. Our method is then applied and we plot the mean size $N_+$ of the augmented graphs and the standard deviation of the sample.The power-law fit suggests  a sub-linear growth of the required additional atoms.}
    \label{fig:overhead_ancillary}
\end{figure}

  
The procedure described above enables us to replicate the connectivity of any input graph G of maximum degree six in three dimensions, at the expense of adding ancillary vertices. In order to assess the overhead incurred by this embedding, we test the procedure of general graphs of maximum degree 6. For each size, 20 Erdos-Renyi graphs are generated and the size $N_+$ of the augmented graph is recorded. Our simulations seem to indicate a sub-linear overhead in the number of ancillary atoms, as illustrated in Figure\,\ref{fig:overhead_ancillary} where we show the number of vertices in the augmented graph $N_+$ with respect to the size of the original graph $N$. Our method would be impractical if the number of ancillary atoms exploded or if the size of the edges grew exponentially with respect to the graph size. Indeed, the Lieb-Robinson bounds\,\cite{lieb1972finite} would imply that information could not propagate efficiently through the ancillary paths. Luckily, the necessary volume to draw the OGD of a graph was proven to be polynomially bounded\,\cite{Eades1996}. Precisely, every $N$-vertex degree-6 graph admits an OGD in $\mathcal{O}(N^{3/2})$ volume and that bound is best possible for some degree-6 graphs. The authors give an explicit algorithm that places all vertices on a $\mathcal{O}(N) \times \mathcal{O}(N)$ grid in a $2D$ plane and draws each edge with at most $16$ bends. The growing number of atoms that can be experimentally trapped in recent experimental setups\,\cite{Schymik_2022} is an encouraging sign that the overhead in the number of atoms required in our method is reasonable.\\

\section{Conclusion}
We described here the versatility of neutral atom platforms at solving hard graph problems. 
Previous implementations of MIS solutions using atoms trapped in optical tweezers were either limited to UD graphs\,\cite{Dalyac2021, queraMIS}, or using a large number of ancillary qubits\,\cite{quera_gadget}.
We have successfully illustrated that this approach can be extended to a larger class of graphs, by solving the MIS problem on a minimal non-unit disk {\it hard graph} using a 3D array of atoms. 
The results showcase the validity of the approach for unit-ball graphs.

Furthermore, we described a method embedding the MIS problem over non-local graphs as Unit-Ball graphs in 3D space.
This procedure is guaranteed to run in a time growing polynomially with the input graph size, and can be used, in particular, to encode bounded-degree graphs of maximum degree 6, for which no PTAS exists unless $P = NP$. 
As neutral atom platforms develop and benefit from additional hardware features, we expect our method to become more and more relevant.
In particular, the validity of the OGD relies on the ancillary chains being in an antiferromagnetic state, which can be enforced using local detunings. 
A very intriguing and exciting prospect is to qualify both theoretically and experimentally the capabilities of quantum devices in approximately solving the MIS on this classes of hard graphs like the bounded-degree graphs.
We believe that understanding from a theoretical perspective the capabilities of quantum approaches in finding guaranteed performance ratio for NP-complete problems is an important question in the quest for practical quantum advantage.


\section*{Acknowledgments}
We thank Marc Porcheron, Erik Jan van Leeuwen, Lucas Leclerc, Alex B. Grilo, Elham Kashefi, Thierry Lahaye and Antoine Browaeys for thoughtful discussions and remarks. We acknowledge support from the region Ile-de-France through the AQUARE project, as part of the PAQ program.

\appendix



\section{Optimal local detuning on ancillary atoms}
\label{appendix_local_detuning}

We estimate in this proof the lower and upper bound for the local detuning on the two ancillary atoms of an augmented edge. Let $E_{i_1i_3i_4i_2}$ be the energy associated to the bit-string $i_1i_3i_4i_2$, where $i_k \in \{0,1\}$ and $k$ is the label of the atom ($1,2$ are the main atoms, and $3, 4$ the ancillas. We want to ensure the following inequalities: 
\begin{equation}
    \begin{cases} 
    E_{1010} < E_{1001} \\ 
    E_{1010} < E_{0110} \\
    \end{cases}
\end{equation}

We show calculations in the case $\delta_i = 0$ for original atoms (1 and 2). We therefore have that 

\begin{equation}
\begin{split}
    &\begin{cases} 
    -2\delta - \delta_i + \frac{1}{2^6}U < -2\delta + \frac{1}{3^6}U  \\ 
    -2\delta - \delta_i + \frac{1}{2^6}U  < -2 (\delta + \delta_i) + U \\
    \end{cases} \\
    &\iff \delta_i \in \left[\left( \frac{1}{2^6} - \frac{1}{3^6}\right) \times U, \left(1 - \frac{1}{2^6}\right) \times U\right]
\end{split}
\end{equation}

Taking $\delta_i = U/2$ for all ancillary atoms is a safe spot.
\bibliographystyle{unsrtnat}
\bibliography{refs}

\end{document}